\documentclass[aps,11pt,a4paper, reprint]{iopart}
\usepackage[english]{babel}
\usepackage[utf8]{inputenc}
\expandafter\let\csname equation*\endcsname\relax

\expandafter\let\csname endequation*\endcsname\relax

\usepackage{amsmath}
\usepackage{graphicx}
\usepackage{tabularx}
\usepackage{geometry}
\usepackage{float}
\usepackage{geometry}
\usepackage{parskip}
\usepackage{xcolor}
\usepackage{hyperref}

\DeclareMathOperator{\sinc}{sinc}

\begin{document}

\title{Flexible source of correlated photons based on LNOI rib waveguides}
\author{L. Ebers$^{1}$\footnote[7]{These authors contributed equally to this work. \label{note}}, A. Ferreri$^{2,3}$\footnotemark[1], M. Hammer$^1$, M. Albert$^2$, C. Meier$^2$, J. Förstner$^1$, P. R. Sharapova$^2$}
\address{$^1$ Paderborn University, Theoretical Electrical Engineering and CeOPP, Warburger Stra\ss{}e 100, D-33098 Paderborn, Germany}
\address{$^2$ Paderborn University, Department of Physics, Warburger Stra\ss{}e 100, D-33098 Paderborn, Germany}
\address{$^3$ Institute for Quantum Computing Analytics (PGI-12), Forschungszentrum Jülich, 52425 Jülich, Germany}

\begin{abstract}
	Lithium niobate on insulator (LNOI) has a great potential for photonic integrated circuits, providing substantial versatility in design of various integrated components. To properly use these components in the implementation of different quantum protocols, photons with different properties are required. In this paper, we theoretically demonstrate a flexible source of correlated photons built on the LNOI waveguide of a special geometry. This source is based on the parametric down-conversion (PDC) process, in which the signal and idler photons are generated at the telecom wavelength and have different spatial profiles and polarizations, but the same group velocities. Such features facilitate electro-optical manipulations with photons and, at the same time, do not require an additional compensation for the time delay. We show how the spectral properties of the generated photons and the number of their frequency modes can be controlled depending on the pump characteristics and the waveguide length. Finally, we discuss a special regime, in which narrowband light with strong frequency correlations is generated at the telecom wavelength.
\end{abstract}

\section{Introduction}
Lithium Niobate on Insulator (LNOI) combines excellent nonlinear and electro-optical properties of bulk lithium niobate with the design freedom and sophisticated fabrication procedures known for silicon photonics \cite{Qi20,Boe18, Wei16}, and is therefore a promising platform for integrated quantum photonic circuits \cite{Wan20,Sar21} and neuromorphic photonics \cite{Sha21}. The current progress in fabrication and domain engineering in LNOI \cite{Sun20} allows to produce waveguides of various geometries, poling and optical properties, which makes this material very suitable for efficient second-order processes, such as second harmonic generation \cite{Luo18} and parametric down-conversion (PDC) \cite{Ma20}.

Parametric down-conversion is one of the most powerful processes for producing entangled photons \cite{Kur01, San21}, which are an essential part of quantum optical devices and circuits \cite{Zei17}. For quantum communication networks \cite{Gis02}, compact sources of entangled narrowband photons play a significant role, since they can reduce mode chromatic dispersion and transmission loss over long distances \cite{Mon14}, as well as improve the quantum light-matter interaction \cite{Pra19}. It has recently been shown that  LNOI is a promising material for generating entangled photons with a high rate \cite{Xue21} and for creating the directional pairs of entangled photons \cite{Dua20}.

In lithium niobate, the type-II PDC process is a more desirable source of correlated photons since it involves the second strongest component of the nonlinear susceptibility tensor $d_{31}$ and, at the same time, produces orthogonally polarized photons which can be easily manipulated by electro-optical tools \cite{Sha17}. However, due to their different polarizations, signal and idler photons acquire a certain time delay when leaving a piece of lithium niobate. While in bulk systems this delay can be easily compensated for, integrated devices require additional complexity to neutralize it. This situation can be improved in thin films of LNOI, where, due to a specific geometry, the signal and idler photons can posses the same effective refractive indices and propagate together, having different polarizations, spatial profiles and frequency distributions. 

In this work, we study LNOI rib waveguides of a certain geometry, which results in the same effective refractive indices of the orthogonally polarized signal and idler PDC photons produced at the telecom wavelength. The described PDC photons are generated in different spatial modes (TM$_0$ and TE$_2$) and, therefore, exhibit entanglement in both polarization and spatial domains. We investigate the Z-cut and X-cut configurations of such a waveguide and show how the spectral properties of the generated PDC photons can be modified by varying the pump characteristics and the waveguide length. We demonstrate the generation of multimode PDC light with strong frequency correlations as well as decorrelated photons in a single frequency mode. Finally, we discuss regimes in which entangled photons with narrowband spectral profiles are created inside the presented waveguide.

\section{Nonlinear interaction of modes in LNOI rib waveguides}

\subsection{LNOI rib waveguide structure}
\begin{figure}[H]
	\centering
	\includegraphics[width = .6\textwidth]{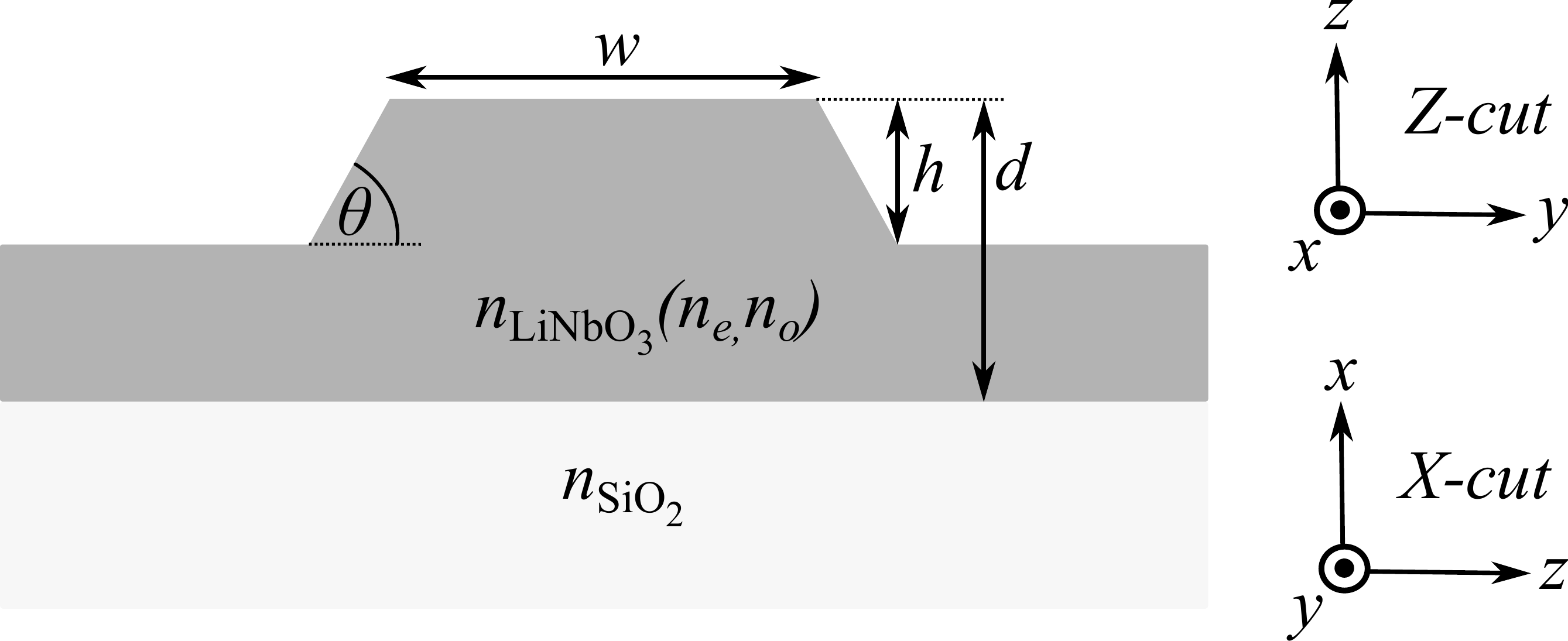}
	\caption{A 2D sketch of the LNOI rib waveguide of width $w$, height $d$, etch depth $h$ and angle $\theta=45^\circ$. Different coordinate systems show the profile of the waveguide in X- and Z-cut configurations. See Tab.~\ref{RI} and Fig.~\ref{scan_f} for the corresponding refractive index values.}
	\label{sketch}
\end{figure}

In this work, we investigate the frequency degenerate type-II PDC process which takes place in the LNOI rib waveguide and results in the signal and idler photons at the telecom wavelength $\lambda_s=\lambda_i=1.55\,\mu\text{m}$, while the wavelength of the pump is $\lambda_p=0.775\,\mu\text{m}$. A general 2D sketch of the studied waveguide is shown in Fig.~\ref{sketch} with either X-cut ($y$ is the propagation direction) or Z-cut ($x$ is the propagation direction) coordinate system. The structure is assumed to be constant along the propagation direction and the optical axis of the anisotropic lithium niobate crystal is always along the $z$-direction. The geometry parameters are given by the width $w$, the height $d$, the etch depth $h$ and the angle $\theta=45^\circ$. 

\begin{figure}[H]
	\centering
	\includegraphics[width = .9\textwidth]{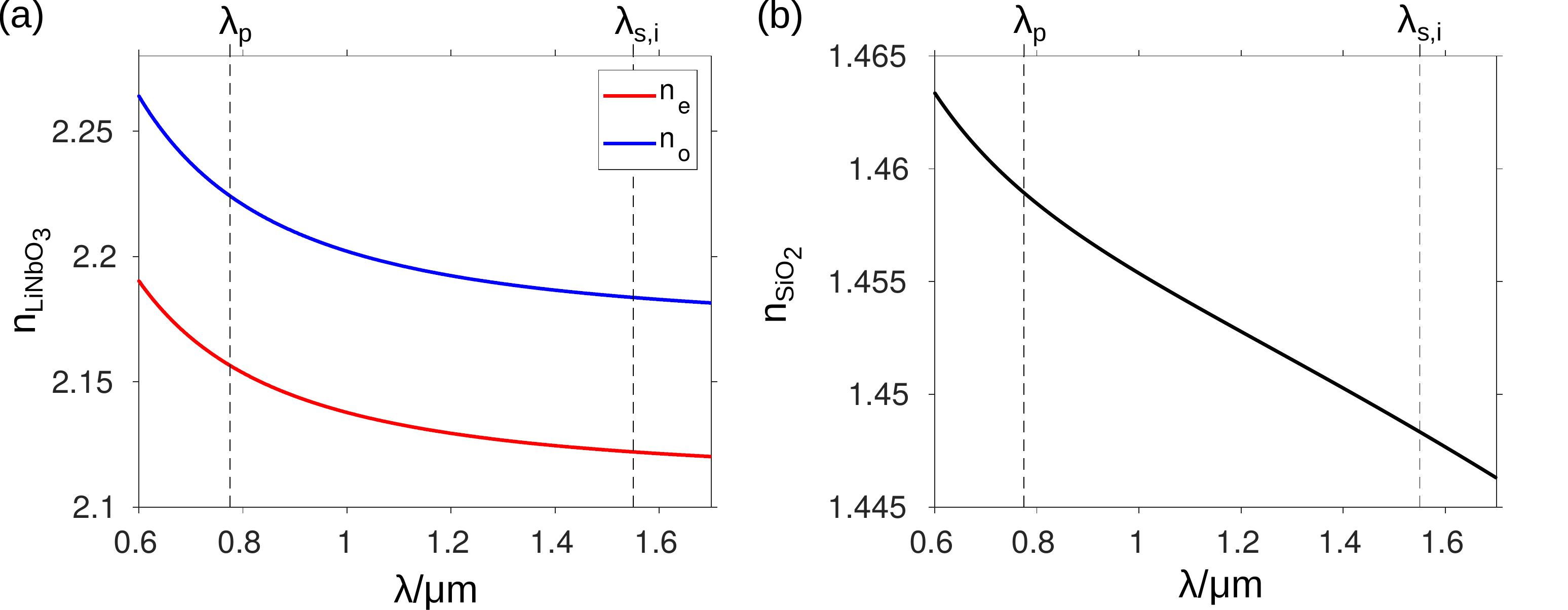}
	\caption{Wavelength dependence of the refractive indices for (a) lithium niobate (blue line - ordinary $n_o$ and red line - extraordinary $n_e$ refractive indices) and (b) silicon dioxide. The corresponding signal/idler $\lambda_{s,i}$ and pump $\lambda_p$ wavelengths from Tab.~\ref{RI} are marked by the vertical dashed lines. }
	\label{scan_f}
\end{figure}

The corresponding refractive indices for both relevant wavelengths in bulk $\mathrm{LiNbO_3}$ and $\mathrm{SiO_2}$ are displayed in Tab.~\ref{RI}, while the respective wavelength scans are shown in Fig.~\ref{scan_f}. The values for the silicon refractive indices are adopted from \cite{Her98}. The values for the lithium niobate refractive indices were measured experimentally using spectroscopic ellipsometry and then were fitted using a Tauc-Lorentz oscillator-based model (see Appendix C).

\begin{table}[H]
	\centering
	\begin{tabular}{l|c|c|c|}
		& $\mathrm{SiO_2}$	& \multicolumn{2}{c|}{$\mathrm{LiNbO_3}$}\\
		& $n$ & $n_{e}$ & $n_o$\\
		\hline
		$\lambda_p=0.775\,\mu\text{m}$ & 1.4589 & 2.1565 & 2.2242\\
		$\lambda_{s,i}=1.55\,\mu\text{m}$ & 1.4483 & 2.122 & 2.1837
	\end{tabular}
	\caption{Refractive indices of the signal/idler and the pump photons in bulk $\mathrm{LiNbO_3}$ and $\mathrm{SiO_2}$.}
	\label{RI}
\end{table}

\subsection{Spatial mode profiles}

\begin{figure}
	\centering
	\includegraphics[width=.9\textwidth]{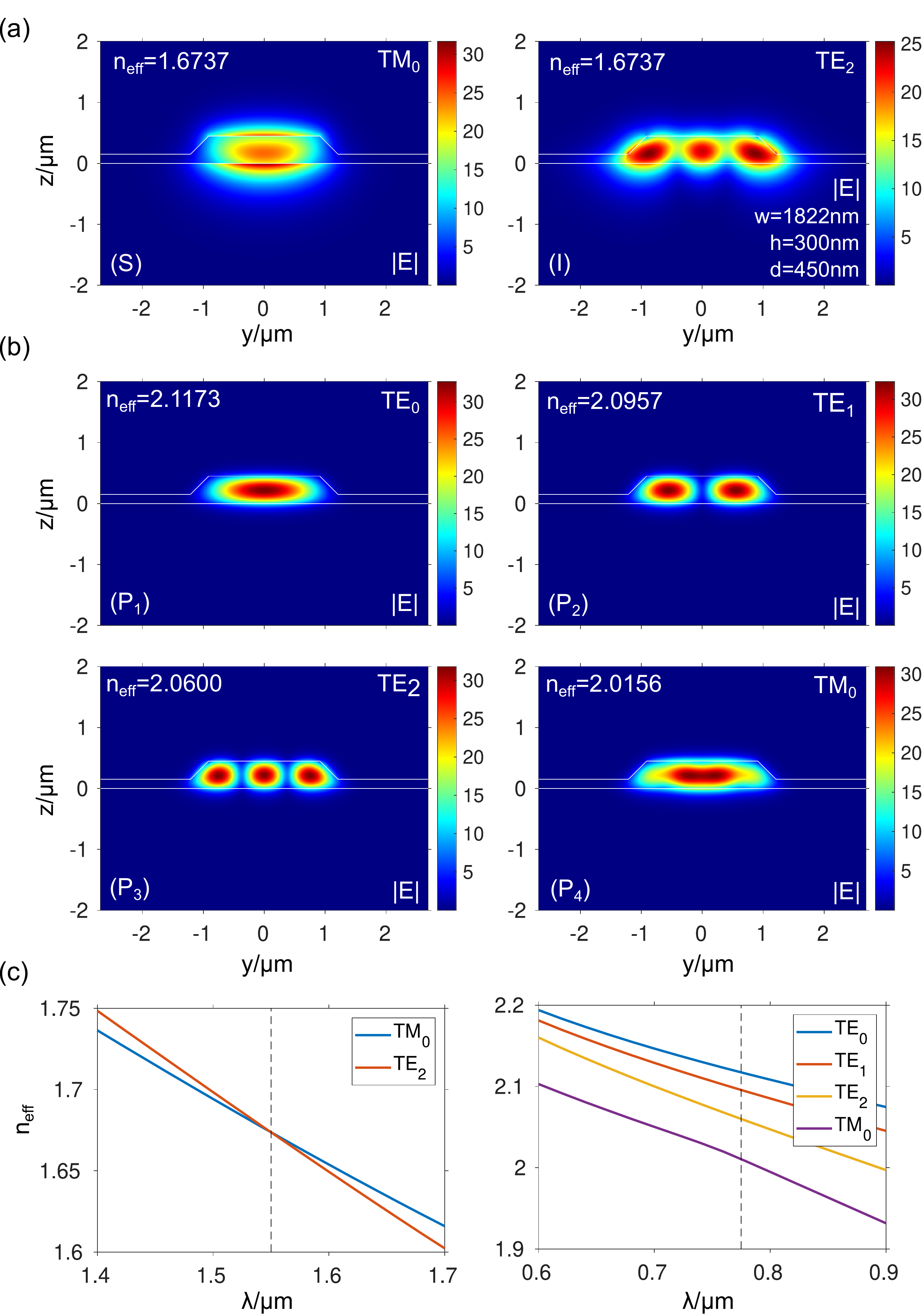}
	\caption{Z-cut structure with geometry parameters $w=2000\,\text{nm}$, $h=300\,\text{nm}$, $d=450\,\text{nm}$. Plots of the absolute electric field $|\mathbf{E}|$ (in V/$\mu$m) for signal (S) and idler (I) modes in (a) and possible pump modes (P$_1$)-(P$_4$) in (b). In (c) the effective refractive index $n_\mathrm{eff}$ versus the wavelength $\lambda$ is shown for signal and idler modes (left plot) and possible pump modes (right plot). All modes are power normalized, i.e. $\int S_x\mathrm{d}z \mathrm{d}y =1\,\text{W}$, where $S_x$ is the Poynting vector in the $x$-direction.}
	\label{Z-Cut} 
\end{figure}

\begin{figure}
	\centering
	\includegraphics[width=.9\textwidth]{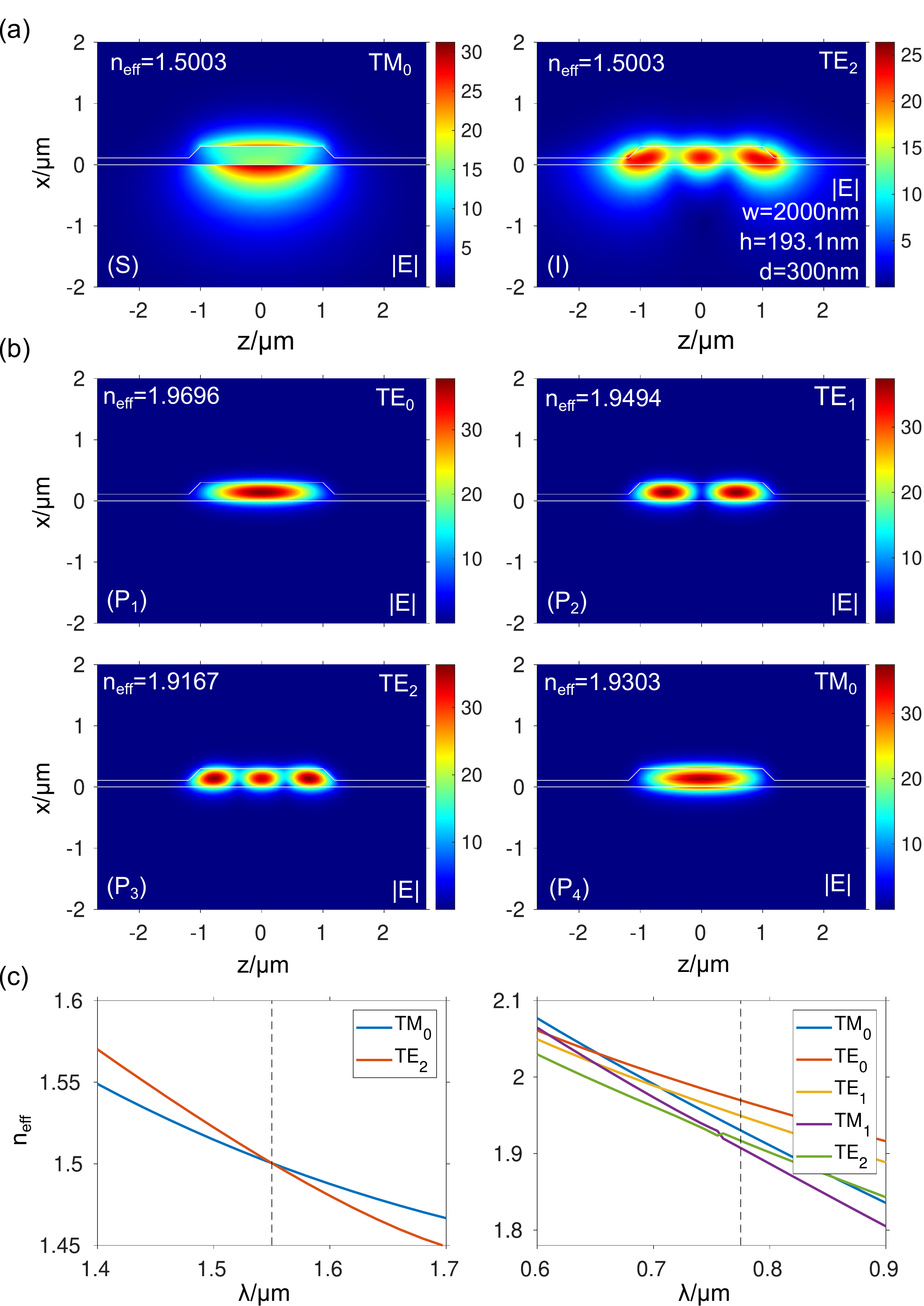}
	\caption{X-cut structure with geometry parameters $w=2000\,\text{nm}$, $h=193.1\,\text{nm}$, $d=300\,\text{nm}$. Plots of the absolute electric field $|\mathbf{E}|$ (in V/$\mu$m) for signal (S) and idler (I) modes in (a) and possible pump modes (P$_1$)-(P$_4$) in (b). In (c) the effective refractive index $n_\mathrm{eff}$ versus the wavelength $\lambda$ is shown for signal and idler modes (left plot) and possible pump modes (right plot; additionally, the TM$_1$ mode is displayed to explain the jump in the curve of the TE$_2$ mode due to the crossing of the graphs). All modes are power normalized, i.e. $\int S_y \mathrm{d}z \mathrm{d}x=1\,\text{W}$, where $S_y$ is the Poynting vector in the $y$-direction.}
	\label{X-Cut}
\end{figure}

To find a suitable waveguide geometry in which the signal and idler photons have the same effective refractive indices while being orthogonally polarized, 
we made use of the finite element solver COMSOL Multiphysics \cite{comsol} and calculate the guided modes (the degenerated signal/idler and pump modes) using mode analysis studies. For simplicity, the search for suitable modes is carried out in reverse to the PDC process itself: first, we look for degenerate signal and idler modes at $\lambda=1.55\,\mu\text{m}$ by varying the geometry parameters $w,d,h$ of the waveguide. The resulting geometry then determines the corresponding variety of possible pump modes at $\lambda=0.775\,\mu\text{m}$. The absolute electric field $|\mathbf{E}|$ distribution of the signal (TM$_0$) and idler (TE$_2$) photons together with the absolute electric field of several pump modes supported by the waveguide are presented in Fig.~\ref{Z-Cut} (a,b) and Fig.~\ref{X-Cut} (a,b) for the Z-cut and X-cut configurations, respectively.

The presence of the thick silica buffer layer modifies the refractive indices of photons propagating in a thin lithium niobate film, while the geometry of the waveguide makes it possible to equalize the refractive indices of various spatial modes at a fixed wavelength. In the left panels of Fig.~\ref{Z-Cut} (c) and Fig.~\ref{X-Cut} (c), the effective refractive indices of the signal and idler photons are presented as a function of wavelength for the Z-cut and X-cut configurations, respectively. One can observe that at the a telecom wavelength $\lambda=1.55\,\mu\text{m}$, the TM$_0$ and TE$_2$ modes have the same effective refractive indices ($n_\mathrm{eff}=1.6737$ for the Z-cut and $n_\mathrm{eff}=1.5003$ for the X-cut configuration), and, therefore, the same propagation constants. This means that with such a PDC source, no additional time delay compensation between the signal and idler photons is required, which can significantly simplify the design of integrated circuits. Moreover, since the signal and idler photons are generated in different spatial modes, they become entangled in the spatial degrees of freedom. In the right panels of Fig.~\ref{Z-Cut} (c) and Fig.~\ref{X-Cut} (c), the wavelength scan of the refractive index of the pump photon is shown for different supported pump modes. In the X-cut configuration (Fig.~\ref{X-Cut} (c)), there are several intersection points for which the refractive indices of different pump modes coincide. At these wavelengths, two parallel PDC processes originating from different pump modes can be phasematched simultaneously, that can be used to create multipartite entanglement. However, since there is no such kind of intersections for the target pump wavelength $\lambda_p=0.775\,\mu\text{m}$ in the current waveguide geometry, we will not focus on this effect in the later discussion.

Classically, the nonlinear interaction of the optical fields is described by the coupled amplitude equations (Appendix A, Eq.~(\ref{A}); depending on the orientation of the crystal, the propagation direction has to be adapted properly, c.f Fig.~\ref{sketch}). Such equations allow us to determine the coupling strength between different modes using coupling coefficients $\kappa_s, \kappa_i, \kappa_p$ (Appendix A, Eq.~(ref{C})). When considering such coupling coefficients, the relevant area for integration is limited to the lithium niobate layer, which only has a nonzero second-order susceptibility $\chi^{(2)}$. In simulations, we assume that a $\chi^{(2)}$-tensor is related to the following nonzero tensor components $d_{31}=4.6\,\frac{\text{pm}}{\text{V}}$, $d_{33}=25\,\frac{\text{pm}}{\text{V}}$ and $d_{22}=2.2\,\frac{\text{pm}}{\text{V}}$ adopted from \cite{Wan18a,Sho97}. Furthermore, due to the symmetry of the $\chi^{(2)}$-tensor \cite{Boy08}, we can identify the equality $\kappa_s=\kappa_i=\frac{1}{2}\kappa_p^*=:\kappa$. Therefore, in the further discussion, we will indicate only the values for $\kappa$-coefficients. 



\subsection{Coupling strength of the PDC process}

In the spontaneous regime, the PDC wavefunction can be found using the first-order perturbation theory (see Appendix B):
\begin{equation}
	|\psi_{\mathrm{PDC}}\rangle \simeq \Gamma \int \mathrm{d}\omega_s \mathrm{d}\omega_i F_N(\omega_s,\omega_i)\hat a^\dagger(\omega_s)\hat a^\dagger(\omega_i)|0\rangle,
	\label{statepdc2}
\end{equation}
where $\Gamma$ is the coupling strength which depends on the pump power, the waveguide length $L$, as well as on the overlap of the spatial profiles of the pump, signal and idler photons in the waveguide, see Appendix B, while 
\begin{equation}
	F_N(\omega_s,\omega_i)=\frac{1}{N_F} \mathrm{e}^{-\frac{(\omega_p-\omega_s-\omega_i)^2\tau^2}{2}}\sinc\left[\frac{\Delta \beta L}{2}\right]\mathrm{e}^{\mathrm{i}\frac{\Delta \beta L}{2}},
	\label{tpa}
\end{equation}
is the normalized joint spectral amplitude (JSA) of the process, $N_F$ is the normalization constant which is connected with the bandwidth of the generated light, $\tau$ is the pulse duration of the pump laser, $\omega_{p,s,i}$ and $k_{p,s,i}=n_{\mathrm{eff}, p, s,i} \frac{\omega_{p,s,i}}{c} $ are the frequencies and wavevectors of the pump, signal and idler photons, respectively, $c$ is the speed of light. For the configurations considered in this work, the phase matching $\Delta k = k_p-k_s-k_i=0$ is nowhere fulfilled, therefore, to suppress an oscillating behavior of the amplitudes, a poling of the waveguide with a period $\Lambda$ is required. Such a poling results in a quasi-phase matching $\Delta \beta =\Delta k - \frac{2 \pi }{\Lambda}$. The first exponential term in Eq.~(\ref{tpa}) includes the pump pulse duration and is usually called the pump function, while the sinc-term determines the phasematching function that reflects the properties of the material.

The coupling strength $\Gamma$ determines the number of generated photon pairs, which increases with increasing pump power and waveguide length. Therefore, to estimate the relative strength of the various processes under consideration, it is more convenient to introduce the coupling strength per unit length, pump power $P$, time and spectral range, namely, $ |\Gamma| / (L \sqrt{P} \tau N_F) \approx |\kappa|/ 3 $, see Appendix B, where $\kappa$ is the coupling coefficient discussed early and presented in Appendix A. 

The calculated $|\kappa|$-coefficients for different pump modes (TE$_0$-TE$_2$ and TM$_0$) are displayed in Tab.~\ref{Tab} for X-cut and Z-cut configurations. In both systems, the appropriate frequency-degenerate signal and idler photons are created in the TM$_0$ and TE$_2$ mode, respectively. Additionally, the required poling period $\Lambda$ is shown in the same table. It can be seen that the greatest value of $|\kappa|$ is achieved by the Z-cut configuration for the TE$_2$ pump mode with a magnitude of $220.40\,\text{W}^{-1/2}\text{m}^{-1}$, which results in a brightness of ${5\times 10^6~ \text{pairs} / (\text{s}\cdot \text{mW} \cdot \text{GHz})}$ for a $30\,\text{mm}$ long waveguide pumped by a continuous wave laser. This brightness is about of ten times larger than the value reported in \cite{Fuj07} for a lithium niobate waveguide of the same length and is comparable with values obtained in \cite{Xue21}.



\begin{table}[H]
	\centering
	\begin{tabular}{c||c|c|c|c||c|c|c|c}
		& \multicolumn{4}{c||}{X-cut}& \multicolumn{4}{c}{Z-cut}\\
		pump & TE$_0$ & TE$_1$ & TE$_2$ & TM$_0$ & TE$_0$ & TE$_1$ & TE$_2$ & TM$_0$\\
		\hline
		$|\kappa|[\text{W}^{-1/2}\text{m}^{-1}]$ & 11.54 & 34.87 & 7.44 & 40.15 & 65.91 & 9.50 & 220.40 & 1.86\\
		$\Lambda[\mu\,\text{m}]$ & 1.65 & 1.73 & 1.86 & 1.80 & 1.75 & 1.84 & 2.01 & 2.27
	\end{tabular}
	\caption{Calculated $|\kappa|$-coefficients and poling periods $\Lambda$ for different pump modes (TE$_0$-TE$_2$, TM$_0$) at the wavelength $\lambda=0.775\,\mu\text{m}$ for the X-cut and Z-cut configurations. The signal and idler photons are in the TM$_0$ and TE$_2$ mode, respectively, at $\lambda=1.55\,\mu\text{m}$. The corresponding field plots are illustrated in Fig.~\ref{Z-Cut} (a,b) and Fig.~\ref{X-Cut} (a,b).}
	\label{Tab}
\end{table}

\section{Quantum properties of the generated light}

The effective refractive index of modes and, therefore, the spectral properties of the generated light strongly depend on the geometry of the waveguide. In this section, we investigate the spectral properties of PDC light for the geometries discussed above when the signal (TM$_0$) and idler (TE$_0$) photons have the same effective refractive indices.

\subsection{Z-cut configuration}

\begin{figure}[H]
	\includegraphics[width=1\linewidth]{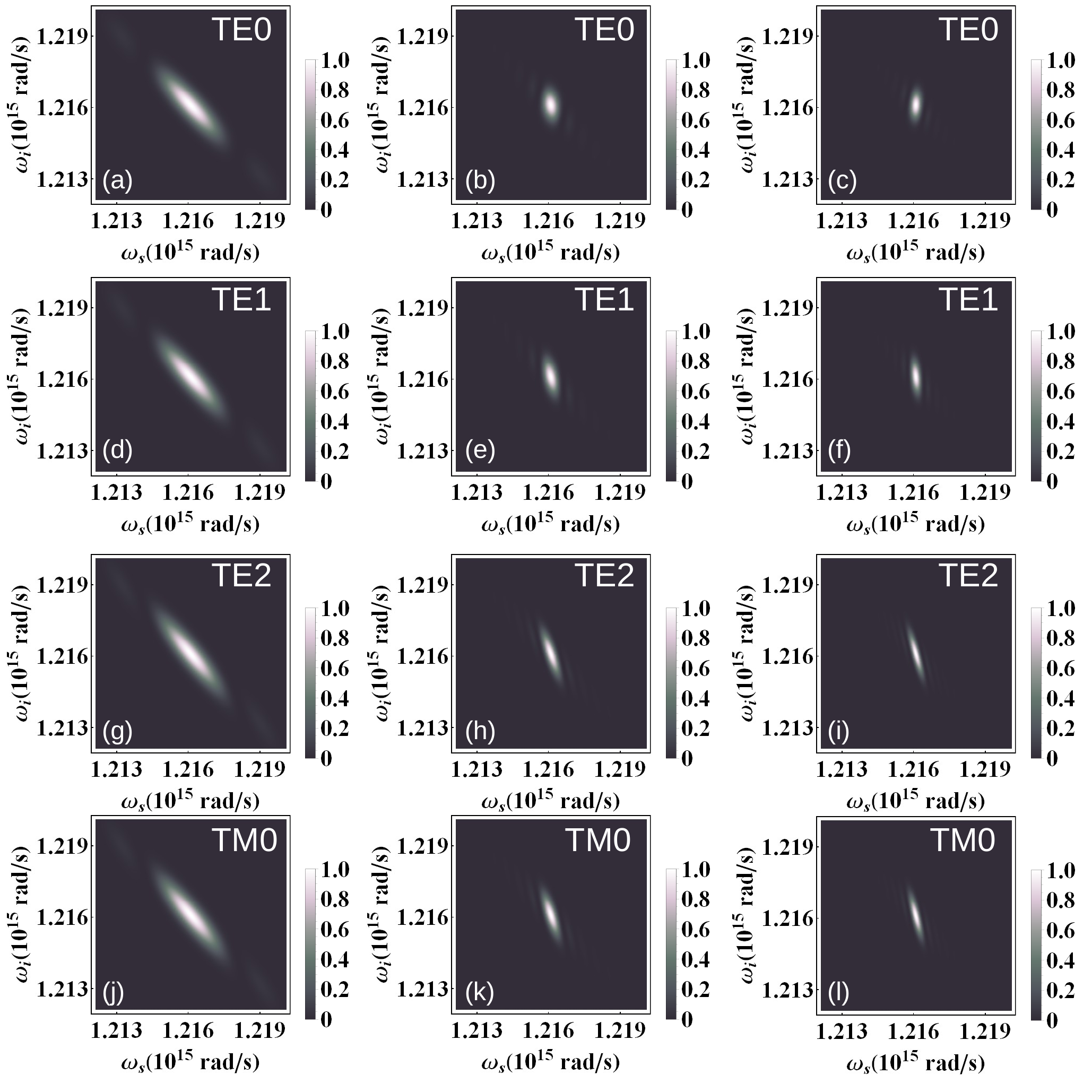}
	\caption{Z-cut structure. Joint spectral intensities for various pump modes and crystal lengths $L = 0.7$\,cm (left column), $L = 2.6$\,cm (middle column), and $L=4$\,cm (right column) with a pulse duration $\tau=1.7$\,ps.}
	\label{cut_Z_TPI}
\end{figure}

For the fixed modes of the signal and idler photons, the pump mode plays an important role, since different pump modes result in different phasematching functions and, therefore, in different joint spectral amplitudes of the PDC process. In Fig.~\ref{cut_Z_TPI}, the joint spectral intensities (JSI) $I(\omega_s, \omega_i)=|F_N(\omega_s, \omega_i)|^2$ of the generated light are shown for different pump modes and crystal lengths in the case of a pulsed pump with the pulse duration of $\tau=1.7$\,ps.

According to Eq.(\ref{tpa}), the smaller the length of the crystal is, the broader the phasematching function becomes. At the same time, the pump function cuts out a region of a certain width around the antidiagonal. If the width of the phasematching function is broader than the width of the pump function, the profile of the JSI is mainly determined by the properties of the pump. Therefore, the resulting joint spectral intensity is represented by an ellipse oriented to the left (left column in Fig.~\ref{cut_Z_TPI}). In this case, the joint spectral intensity is not sensitive to the phasematching function, which leads to almost the same JSI shape for all pump modes.

For a fixed pulse duration, with an increase in the crystal length the properties of the phasematching function become more pronounced. If the width of the phasematching function is narrower than the width of the pump function, the shape of the JSI is mainly determined by the phasematching function profile, that can be seen in the right column of Fig.~\ref{cut_Z_TPI}. In this case, the TE$_1$ pump mode results in an ellipse that is almost perpendicular to the $y$-axis, while the TE$_0$ pump mode leads to the right-oriented ellipse in the JSI. At the same time, the TM$_0$ and TE$_2$ pump modes result in joint spectral intensities with an orientation similar to that of the bulk lithium niobate.

The change in the orientation of JSI with the crystal length causes the JSI to have a nearly circular shape for an appropriate set of parameters (Fig.~\ref{cut_Z_TPI} (b)), which indicates an almost frequency-decorrelated state resulting in a frequency-single-mode regime. Indeed, the effective number of spectral modes can be estimated using the Schmidt number $K=1/ \sum_n \lambda_n^2$, where $\lambda_n$ are the eigenvalues of the Schmidt decomposition of the normalized joint spectral amplitude:
\begin{equation}
	F_N(\omega_s,\omega_i)= \sum_n \sqrt{\lambda_n} u_n(\omega_s) v_n(\omega_i),
\end{equation}
$u_n(\omega_s)$ and $v_n(\omega_i)$ are the spectral eigenmodes of the system, which diagonalize the frequency-dependent PDC Hamiltonian \cite{Sha18}.

The Schmidt number for different pump modes and crystal lengths is presented in Tab.~\ref{tab_K_Z}. One can observe that in the case of TE$_1$ and TE$_0$ pump modes, an almost single-mode regime can be achieved. An imperfection of the single-mode regime is caused by the pronounced side peaks in the JSI (see Fig.~\ref{cut_Z_TPI} (b)), which, however, can be filtered with a bandpass filter due to the mostly perpendicular orientation of the JSI. The use of a continuous wave (CW) laser significantly increases the number of spectral modes, resulting in highly frequency-correlated PDC light.
Thus, the number of spectral modes of the described LNOI waveguide can be easily controlled over the entire range from single-mode to highly multimode by changing its length and pump pulse duration. Such a complete control is not possible with bulk lithium niobate due its phasematching function profile. 


\begin{table}[h]
	\centering
	\begin{tabular}{c||c|c|c||c|c|c}
		& \multicolumn{3}{c||}{pulsed pump}& \multicolumn{3}{c}{CW pump}\\
		pump mode & 0.7\,cm & 3\,cm & 4\,cm & 0.7\,cm & 3\,cm & 4\,cm\\
		\hline
		TE$_0$ &  2.14 &  1.15 & 1.19 & 624.38 & 150.18 & 98.73 \\
		TE$_1$ & 2.22 &  1.19 & 1.14 & 624.38 & 150.18 & 98.73 \\
		TE$_2$ & 2.41 &  1.89 & 2.14 & 624.38 & 150.18 & 98.73 \\
		TM$_0$ & 2.37 &  1.68 & 1.84 & 624.38 & 150.18 & 98.73 
	\end{tabular}
	\centering
	\caption{Z-cut structure. The Schmidt number for various pump modes and crystal lengths in the case of a pulsed pump laser with $\tau=1.7$\,ps and a CW laser.}
	\label{tab_K_Z}
\end{table}

\begin{figure}[H]
	\includegraphics[width=1\linewidth]{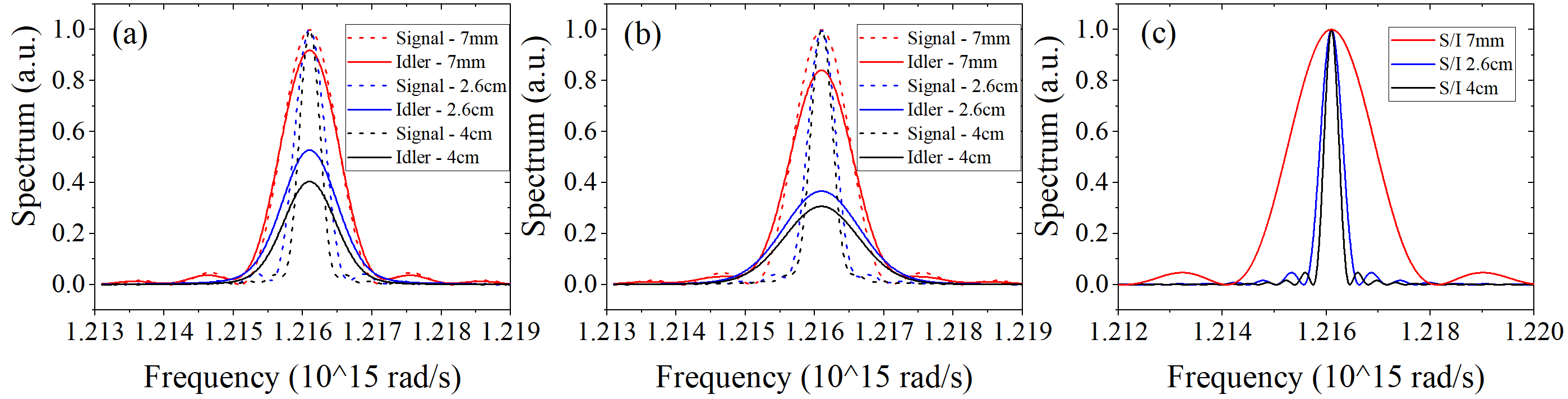}
	\caption{Z-cut structure. Spectra of the signal (dashed lines) and idler (solid lines) photons for various waveguide lengths and pump modes: (a) TE$_0$ pulsed laser ($\tau=1.7$\,ps), (b) TM$_0$ pulsed laser ($\tau=1.7$\,ps) and (c) TE$_0$ CW laser. For the CW laser, the signal and idler spectra coincide.}
	\label{Z_cut_spectra}
\end{figure}

The spectra of the signal and idler photons for the TE$_0$ and TM$_0$ pump modes are presented in Fig.~\ref{Z_cut_spectra}. It can be seen that in the case of a pulsed laser, the frequency profiles of the signal and idler photons are different. This difference becomes more pronounced with the increase in the crystal length, resulting in the simultaneous generation of spectrally narrowband (signal) and broadband (idler) photons. Such photons can be easily separated by a polarizing beam splitter due to their different polarizations, and then can be used separately in quantum circuits. Note that the spectra of the signal and idler photons for the pump mode TE$_2$, which leads to the strongest coupling coefficient $\kappa$, are similar to the spectra for the pump mode TM$_0 $ (Fig.~\ref{Z_cut_spectra} (b)) due to a similar shape of their joint spectral intensities, and therefore are not presented.

The CW laser cuts out very narrow region of the phasematching function around the antidiagonal direction, resulting in the same frequency distributions for the signal and idler photons, see Fig.~\ref{Z_cut_spectra} (c). Herewith, long waveguides lead to the generation of narrowband quantum light with a full width at half maximum FWHM = 1.8\,THz already at a waveguide length of $L=4$\,cm. This is caused by the specific phasematching function of the considered LNOI waveguide and cannot be realized using bulk lithium niobate for the same range of waveguide lengths. Note that, in the case of the CW laser, all pump modes lead to the same spectra of PDC photons. This is why only the spectra for the TE$_0$ pump mode are presented in Fig.~\ref{Z_cut_spectra} (c). At the same time, the CW laser introduces strong frequency correlations into the created PDC light, resulting in the generation of highly correlated narrowband photon pairs.

\subsection{X-cut configuration. }
\begin{figure}[H]
	\includegraphics[width=1\linewidth]{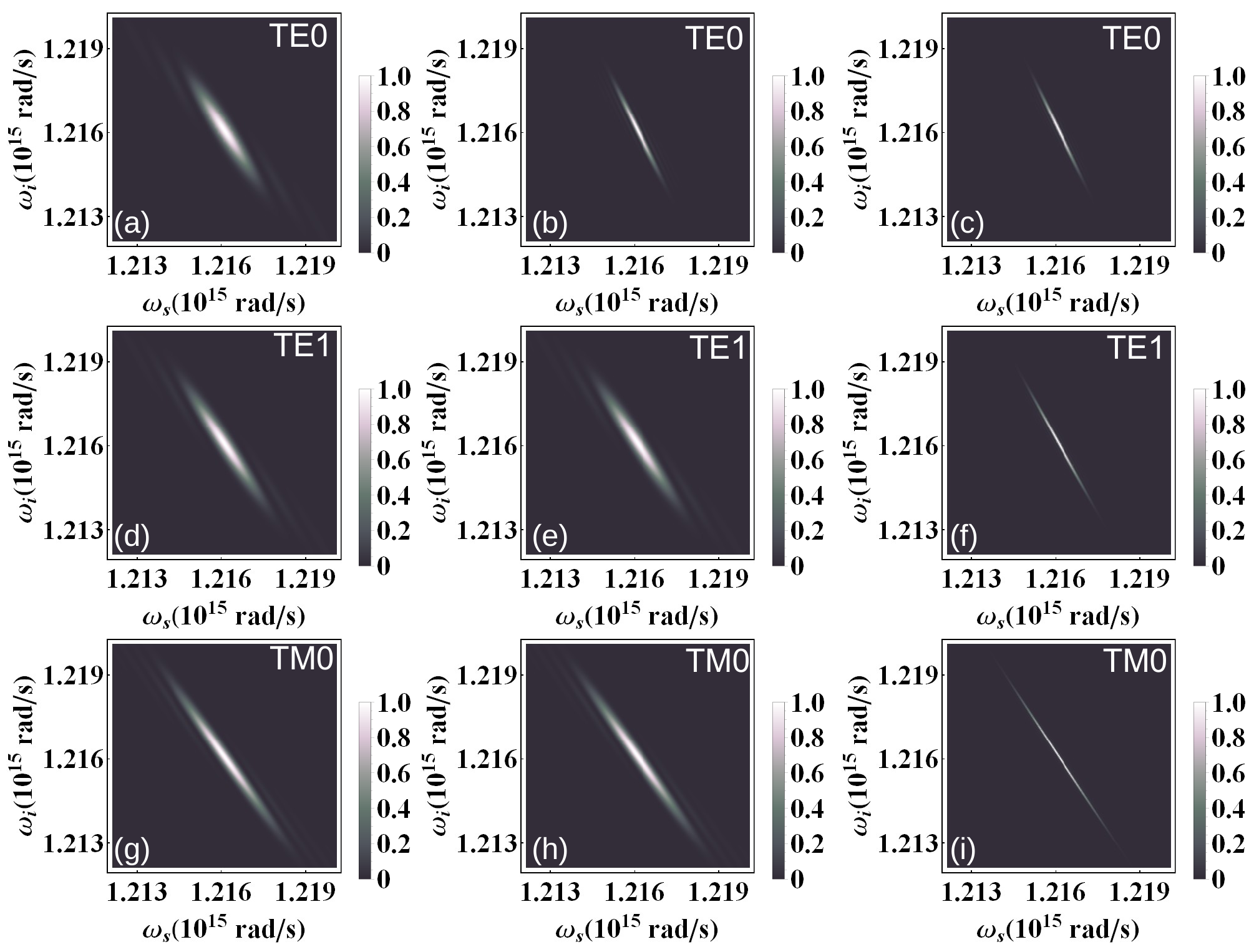}
	\caption{X-cut structure. Joint spectral intensities for various pump modes and crystal lengths $L = 0.7$\,cm (left column), $L = 2.6$\,cm (middle column), and $L=4$\,cm (right column) with a pulse duration $\tau=1.7$\,ps.}
	\label{cut_X_TPI}
\end{figure}

In the X-cut geometry, the phasematching function for all pump modes is much narrower compared to the Z-cut configuration and bulk lithium niobate. Such a property strongly modifies the intensity spectra and makes X-cut configuration more suitable for narrowband light generation. If the spectral width of the pump function is much greater compared to the spectral width of the phasematching function (right column in Fig.~\ref{cut_X_TPI}), the narrowband properties of the phasematching function can be recognized in the JSI plots. Note that in the X-cut configuration, all TE pump modes result in mostly the same phasematching functions, see Fig.~\ref{cut_X_TPI}; the same is valid for the TM pump modes. 

Although in the X-cut configuration, an incline of the phasematching function does not vary with the mode number of the pump as much as in the Z-cut configuration, the regimes with a few number of modes also can be realized in this geometry, see Tab.~\ref{tab_K_X}. Similar to the Z-cut configuration, the number of generated spectral modes can vary from several modes to highly multimode, depending on the crystal length, pump pulse duration, or the spatial profile of the pump mode.

\begin{table}[h]
	\centering
	\begin{tabular}{c||c|c|c||c|c|c}
		& \multicolumn{3}{c||}{pulsed pump}& \multicolumn{3}{c}{CW pump}\\
		pump mode & 0.7 cm & 3 cm & 4 cm & 0.7 cm & 3 cm & 4 cm\\
		\hline
		TE$_0$ & 3.46 & 8.60 & 12.461 & $>$350 & 94.14 & 63.59 \\
		TE$_1$ & 3.82 & 9.99 & 14.53 & $>$350 & 94.14 & 63.59 \\
		TE$_2$ & 4.67 & 13.22 & 19.35 & $>$350 & 94.15 & 63.60 \\
		TM$_0$ & 8.15 & 26.02 & 38.52 & $>$350 & 94.19 & 63.66 
	\end{tabular}
	\caption{X-cut structure. The Schmidt number for various pump modes and crystal lengths in the case of a pulsed pump laser with $\tau=1.7$\,ps and a CW laser.}
	\label{tab_K_X}
\end{table}

As can be seen from Tab.~\ref{tab_K_X}, in the case of a pulsed laser, an increase in the waveguide length leads to an increase in the number of modes. This can also be observed in Fig.~\ref {cut_X_TPI}, where the difference between conditional and unconditional width of JSI \cite{Fed08} increased with increasing the waveguide length. The opposite tendency is observed in the case of continuous pumping and is explained by the off-diagonal orientation of the phasematching function, which leads to the alignment of the conditional and unconditional widths with increasing waveguide length. Finally, the higher the pump mode number is, the larger the number of generated spectral modes becomes.

\begin{figure}[H]
	\includegraphics[width=1\linewidth]{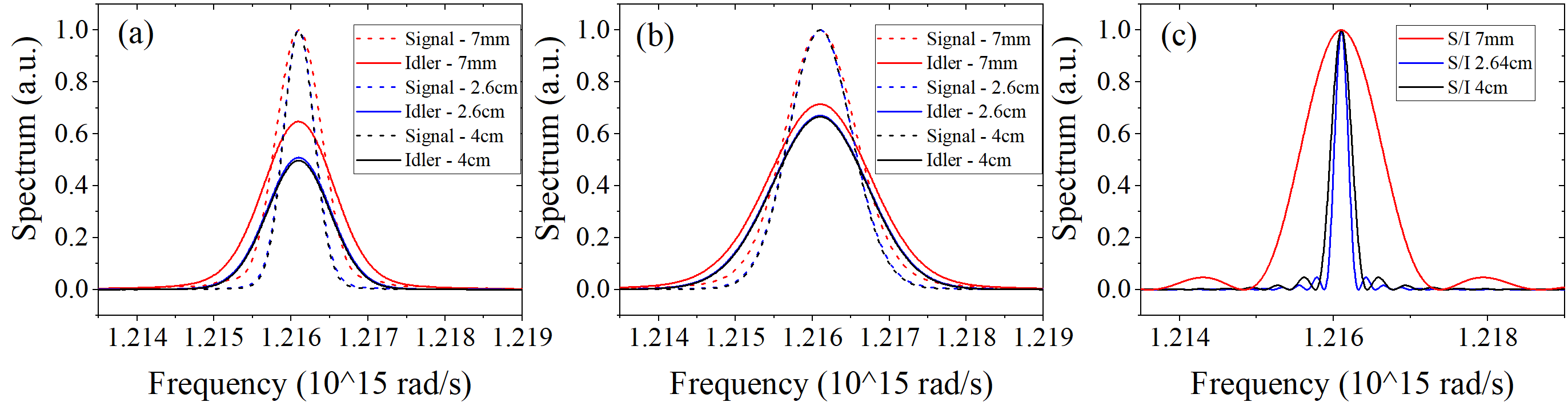}
	\caption{X-cut structure. Spectra of the signal (dashed lines) and idler (solid lines) photons for various waveguide lengths and pump modes: (a) TE0 pulsed laser ($\tau=1.7\text{ps}$), (b) TM0 pulsed laser ($\tau=1.7\text{ps}$) and (c) TE0 CW laser. For the CW laser, the signal and idler spectra coincide.}
	\label{X_cut_spectra}
\end{figure}

Fig.~\ref{X_cut_spectra} presents the spectra of the signal and idler photons. It can be seen that in the case of a pulsed laser, the spectra of both photons change slightly depending on the length of the crystal (especially in the case of the TM$_0$ pump mode), which means that the properties of the generated quantum light are slightly affected by imperfections in the propagation direction. In the CW pump case (Fig.~\ref{X_cut_spectra} (c)), the narrowband pump cuts off the already very narrow phasematching function, resulting in the generation of highly correlated photons with narrowband spectral profiles, where FWHM=1.1\,THz at the length of $L=4$\,cm.

\section{Conclusion}
In this paper, we presented a special geometry of a rib LNOI waveguide, in which the signal and idler PDC photons are generated at a telecom wavelength in modes with different polarizations and spatial profiles, but with the same effective refractive indices and propagation constants. The presented waveguides support various pump modes, which leads to different phasematching functions and spectral properties of the generated light. We investigated the Z-cut and X-cut configurations of such a waveguide and demonstrated how the spectral properties of PDC photons can be controlled by using the waveguide length, pump pulse duration, and varying the spatial profiles of the pump modes. Furthermore, we investigated the mode structure of the generated light and showed that using such a photon source, it is possible to produce the entire range of spectral modes: from a single-mode to highly multimode regimes. In addition, the generation of light at the telecom wavelength with a narrowband terahertz bandwidth and strong frequency correlations in both geometries was theoretically demonstrated and discussed.

\section{Acknowledgements}

Financial support of the Deutsche Forschungsgemeinschaft (DFG) through TRR 142, projects C02 and C05 is gratefully acknowledged. We also acknowledge financial support of the DFG project SH 1228/3-1.

\section*{References}
\bibliographystyle{ieeetr}
\bibliography{references}

\newpage
\appendix
\section{Optical nonlinear interaction of modes}

In optical media, the dielectric polarization $\hat{\mathbf{P}}$ is induced by an electric field $\hat{\mathbf{E}}$. Conventionally, this relationship is mathematically described by a linear dependence. However, when the electric field is substantially strong, non-linear contributions become important and the polarization needs to be expressed through a series expansion ${\hat{\mathbf{P}}=\hat{\mathbf{P}}^{(1)}+\hat{\mathbf{P}}^\mathrm{(NL)}}$ with a linear part $\hat{\mathbf{P}}^{(1)}$ and a remaining non-linear part $\hat{\mathbf{P}}^\mathrm{(NL)}$ \cite{Boy08}. In the following we will briefly derive the equations describing the interaction of waves in nonlinear optical media. For a more detailed description we refer the reader to \cite{Boy08, Cou18, Suh03}.

We consider electric and magnetic fields $(\tilde{\mathbf{E}},\tilde{\mathbf{H}})^T$ in the frequency domain with angular frequencies $\omega=2\pi c/\lambda$, for vacuum wavelengths $\lambda$ and vacuum speed of light $c$,
\begin{equation}
	\begin{pmatrix}
		\hat{\mathbf{E}}\\ \hat{\mathbf{H}}
	\end{pmatrix}(x,y,z,t)=\mathrm{Re}\Bigg\{
	\begin{pmatrix}
		\tilde{\mathbf{E}}\\ \tilde{\mathbf{H}}
	\end{pmatrix}(x,y,z)\mathrm{e}^{\mathrm{i}\omega t}\Bigg\}, 
\end{equation}
which solve the nonlinear Maxwell's equations \cite{Boy08}
\begin{equation} \label{MWE}
	\nabla \times \tilde{\mathbf{E}} = -\mathrm{i}\omega \mu_{0} \tilde{\mathbf{H}}
	\qquad \mathrm{and} \qquad
	\nabla \times \tilde{\mathbf{H}} = \mathrm{i}\omega \epsilon_{0}\epsilon_{r}\tilde{\mathbf{E}}+\mathrm{i}\omega \tilde{\mathbf{P}}^\mathrm{(NL)}.
\end{equation}
In general, for a specific waveguide structure with constant cross section along the propagation direction $z$, we can write the electric and magnetic fields as a complete set of forward and backward propagating waves with propagation constant $k_{n}$ and amplitudes $A_n(z)$, varying with the propagation distance $z$, as
\begin{align} \label{ansatz}
	\begin{pmatrix}
		\tilde{\mathbf{E}}\\
		\tilde{\mathbf{H}}
	\end{pmatrix}(x,y,z)
	= \sum_n \dfrac{A_n(z)}{\sqrt{N_n}} \begin{pmatrix}
		\mathbf{E}_n\\
		\mathbf{H}_n
	\end{pmatrix}(x,y) \mathrm{e}^{\mp \mathrm{i}k_{n} z}.
\end{align}

The couple $(\mathbf{E}_n,\mathbf{H}_n)^T$ identifies the mode profiles in the transverse plane propagating in $\pm z$-direction and can be represented as ${\mathbf{E}_n(x,y)=\mathcal{E}_n \mathbf{E}_{\perp n} (x,y)}$ and ${\mathbf{H}_n(x,y)=\mathcal{H}_n \mathbf{H}_{\perp n} (x,y)}$, where $\mathcal{E}_n$ and $\mathcal{H}_n$ are the amplitudes with the dimension of the electric and magnetic field, respectively, while $\mathbf{E}_{\perp n} (x,y)$ and $\mathbf{H}_{\perp n} (x,y)$ are dimensionless spatial profiles. We introduce a normalization coefficient ${N_n=\frac{1}{4} \int (\mathbf{E}^*_n \times \mathbf{H}_n+\mathbf{E}_n\times \mathbf{H}_n^*)\cdot \mathbf{e}_z \mathrm{d}x\mathrm{d}y}$.

This modal expansion is the fundamental formalism in the coupled mode theory \cite{Suh03}. The amplitudes are then described by the coupled mode theory for interaction of waves in nonlinear optical waveguides \cite{Suh03,Kol04}
\begin{equation}\label{An}
	\partial_z A_n(z)=-\dfrac{\mathrm{i}\omega}{4\sqrt{N_n}} \mathrm{e}^{ \pm \mathrm{i}k_{n} z} \int \mathbf{E}_n^*(x,y)\cdot \tilde{\mathbf{P}}^\mathrm{(NL)}(x,y,z)\mathrm{d}x\mathrm{d}y.
\end{equation}

In this work, we are particularly concentrating on nonlinear optical interactions in media, where the second-order nonlinear susceptibility $\chi^{(2)}$ is nonzero. This allows the creation of electric fields at different frequencies, e.g. parametric down-conversion (PDC) describes a process where a field $\mathbf{{E}}_p$ at frequency $\omega_p$ can split into two new fields $\mathbf{{E}}_s$ and $\mathbf{{E}}_i$ at lower frequencies $\omega_s$ and $\omega_i$ with $\omega_p=\omega_s+\omega_i$ \cite{Cou18}. Here, $\omega_p$ belongs to the pump mode and $\omega_s$ and $\omega_i$ to signal and idler modes. For equation (\ref{An}) this means, we just have to considering three relevant interacting fields for the PDC process, thus $n\in\{p,s,i\}$. We assume that these fields are propagating in the forward $z$-direction (then, only the negative/positive sign is relevant in equation (\ref{ansatz})/(\ref{An})).

Replacing the nonlinear polarization components induced by an electric field in Eq.~(\ref{An}) with the identities \cite{Boy08} 
\begin{align}
	\tilde{\mathbf{P}}^\mathrm{(NL)}_s(x,y,z;\omega_s) &= \frac{2A_i^*A_p\epsilon_0}{\sqrt{N_iN_p}} \chi^{(2)} \mathbf{E}_i^*\mathbf{E}_p\mathrm{e}^{-\mathrm{i}(k_{p}-k_{i})z} \nonumber\\ 
	\tilde{\mathbf{P}}^\mathrm{(NL)}_i(x,y,z;\omega_i)&=\frac{2A_s^*A_p\epsilon_0}{\sqrt{N_sN_p}} \chi^{(2)} \mathbf{E}_s^*\mathbf{E}_p\mathrm{e}^{-\mathrm{i}(k_{p}-k_{s})z}\\
	\tilde{\mathbf{P}}^\mathrm{(NL)}_p(x,y,z;\omega_p)&=\frac{2A_sA_i\epsilon_0}{\sqrt{N_sN_i}} \chi^{(2)} \mathbf{E}_s\mathbf{E}_i \mathrm{e}^{-\mathrm{i}(k_{s}+k_{i})z}\nonumber
\end{align}
results in the following coupled differential equations for the field amplitudes $A_n(z)$, $n\in\{p,s,i\}$
\begin{align}
	\partial_z A_s(z) &=-\dfrac{\mathrm{i}\omega_s \epsilon_0}{2} \frac{A_i^*A_p}{\sqrt{N_sN_iN_p}} \mathrm{e}^{-\mathrm{i}\Delta k z} \int \sum_{l,j,m=x,y,z} \chi^{(2)}_{ljm}E_{s}^{*(l)}E_{i}^{*(j)}{E}_{p}^{(m)} \mathrm{d}x\mathrm{d}y\nonumber\\
	\partial_z A_i(z) &=-\dfrac{\mathrm{i}\omega_i \epsilon_0}{2} \frac{A_s^*A_p}{\sqrt{N_iN_sN_p}} \mathrm{e}^{-\mathrm{i}\Delta k z} \int \sum_{l,j,m=x,y,z} \chi^{(2)}_{ljm}{E}_{i}^{*(l)}{E}_{s}^{*(j)}{E}_{p}^{(m)}\mathrm{d}x\mathrm{d}y\\
	\partial_z A_p(z) &= -\dfrac{\mathrm{i}\omega_p \epsilon_0}{2} \frac{A_sA_i}{\sqrt{N_pN_sN_i}} \mathrm{e}^{\mathrm{i}\Delta k z} \int \sum_{l,j,m=x,y,z} \chi^{(2)}_{ljm}{E}_{p}^{*(l)}{E}_{s}^{(j)} {E}_{i}^{(m)} \mathrm{d}x\mathrm{d}y , \nonumber
\end{align}
where $\Delta k =k_{p}-k_{s}-k_{i}$ is the phase mismatch. Note, that the frequency dependence of $\chi^{(2)}$ is directly neglected as Kleinman's symmetry condition \cite{Boy08} is valid for the processes considered in this work.

For convenience we define the complex values $\kappa_s,\kappa_i,\kappa_p$ via
\begin{align} 
	\kappa_s &= \dfrac{\omega_s\epsilon_0}{2\sqrt{N_sN_iN_p}}\int \sum_{l,j,m=x,y,z}
	\chi^{(2)}_{ljm} E_{s}^{*(l)}E_{i}^{*(j)}{E}_{p}^{(m)} \mathrm{d}x\mathrm{d}y \nonumber \\
	\kappa_i &=\dfrac{\omega_i\epsilon_0}{2\sqrt{N_iN_sN_p}}\int \sum_{l,j,m=x,y,z} 
	\chi^{(2)}_{ljm}{E}_{i}^{*(l)}{E}_{s}^{*(j)}{E}_{p}^{(m)}\mathrm{d}x\mathrm{d}y \label{C}\\
	\kappa_p &=\dfrac{\omega_p\epsilon_0}{2\sqrt{N_pN_sN_i}}\int \sum_{l,j,m=x,y,z} 
	\chi^{(2)}_{ljm}{E}_{p}^{*(l)}{E}_{s}^{(j)} {E}_{i}^{(m)}\mathrm{d}x\mathrm{d}y.\nonumber
\end{align}
The coupled amplitude equations are then simplified as
\begin{align}
	\partial_z A_s(z) &= -\mathrm{i}\kappa_s A_i^*A_p \mathrm{e}^{-\mathrm{i}\Delta k z} \nonumber\\
	\partial_z A_i(z) &= -\mathrm{i}\kappa_i A_s^*A_p \mathrm{e}^{-\mathrm{i}\Delta k z} \label{A}\\
	\partial_z A_p(z) &= -\mathrm{i}\kappa_p A_s A_i \mathrm{e}^{\mathrm{i}\Delta k z} \nonumber.
\end{align}
By assuming a constant-pump approximation \cite{Boy08}, i.e.~$A_p$ is a constant, reduces the equations with $\kappa_s'=\kappa_sA_p$ and $\kappa_i'=\kappa_iA_p$ to 
\begin{align}
	\partial_z A_s(z) &= -\mathrm{i}\kappa_s' A_i^* \mathrm{e}^{-\mathrm{i}\Delta k z}\label{A'}\\
	\partial_z A_i(z) &= -\mathrm{i}\kappa_i' A_s^* \mathrm{e}^{-\mathrm{i}\Delta k z}\nonumber.
\end{align}
Generally, when considering a non-phasematched structure $\Delta k \neq 0$, the solution to these differential equations oscillates with the propagation distance $z$ and the magnitude is mainly determined by the values of $\kappa_s,\kappa_i$. To overcome this oscillating behavior, a quasi-phase matched periodically poled structure can be constructed with poling period $\Lambda=\frac{2\pi}{\Delta k}$.
The parametric down-conversion process describes the creation of signal (s) and idler (i) photons from a photon of the pump (p) using the following Hamiltonian:
\begin{equation}
	\hat H_{\mathrm{PDC}}(t)= \frac{\epsilon_0}{3} \int \mathrm{d^3} r \mathbf{\hat \chi}^{(2)} \hat{\mathbf{E}}_{p}^{(+)}(\vec r,t)\hat{\mathbf{E}}_{s}^{(-)}(\vec r,t)\hat{\mathbf{E}}_{i}^{(-)}(\vec r,t)+ \mathrm{h.c.},
	\label{Hpdc}
\end{equation}
where $\epsilon_0$ is the vacuum permittivity and $ \mathbf{\hat \chi}^{(2)}$ is the nonlinear susceptibility tensor.

The quantized signal (idler) fields are given by:
\begin{equation}
	\hat{\mathbf{E}}_{s}^{(-)}(\vec r,t)= -\mathrm{i} \sum_{\mathbf{k_s}} \sqrt{\frac{\hbar\omega_s}{2\epsilon_0 S_\mathrm{eff} L }} \ \ \hat a_{\mathbf{k_s}}^{\dagger} \mathbf{E}_{\perp s} (x,y) \mathrm{e}^{-\mathrm{i}(\mathbf{k_s} \mathbf{r}-\omega_s t)},
\end{equation}
where $\mathbf{E}_{\perp s} (x,y)$ is the dimensionless transverse mode profile, $L$ is the length of the waveguide, 
$S_\mathrm{eff} = \int \mathrm{d} x \mathrm{d} y \ \ \mathbf{E}_{\perp s}^\ast (x,y) \cdot \epsilon_r (x,y) \mathbf{E}_{\perp s} (x,y)$, and $\epsilon_r (x,y)$ is the tensor of relative permittivity.

Since the transverse size of the considered waveguide structure is rather small compared to the propagation length, in what follows we neglect the transverse wavevectors and take into account the only longitudinal component. Moving from discrete to continuous variables $\sum_k \rightarrow \frac{L}{2 \pi} \int \mathrm{d} k$, making a replacement $a_k \rightarrow \sqrt{\frac{2 \pi}{L}} a(k)$ to preserve the delta-function commutation relation, and taking into account $k= n_\mathrm{eff}(\omega) \omega/c $ \cite{Blo90}, one can obtain the following expression for the signal photon field:
\begin{equation}
	\hat{\mathbf{E}}_{s}^{(-)}(\vec r,t) = - \mathrm{i} \sqrt{\frac{\hbar\omega_{0s} n_\mathrm{eff}(\omega_{0s}) }{4\pi\epsilon_0 c S_\mathrm{eff} }} \int \mathrm{d} \omega_s \ \ \hat a_{s}^{\dagger} (\omega_s) \mathbf{E}_{\perp s} (x,y) \mathrm{e}^{-\mathrm{i}( k_{s} z-\omega_s t)},
	\label{signal_field}
\end{equation}

where we took into account that the center frequency $\omega_{0s}$ is much larger than the frequency distribution of the PDC light, and can be taken out of the integral, $n_\mathrm{eff}(\omega)$ is a slow function with respect to the $\omega$.

A classical pump with the pulse duration $\tau$, the amplitude $\mathcal{E}_{p}$ and the transverse profile $ \mathbf{E}^*_{\perp p} (x,y)$ is given by: 
\begin{equation}
	\hat{\mathbf{E}}_{p}^{(+)}(\vec r,t) = \mathcal{E}_{p} \mathbf{E}^*_{\perp p} (x,y) \mathrm{e}^{-\frac{t^2}{2\tau^2}}\mathrm{e}^{\mathrm{i}(k_p z - \omega_p t)}.
	\label{pump_field}
\end{equation}

A summation over the tensor components $ l, j, m $, each of which can take values $ x, y, z $, leads to the following form of the Hamiltonian:
\begin{equation}
	\hat H_{\mathrm{PDC}}(t)= \frac{\epsilon_0}{3} \int \mathrm{d^3} r \sum_{l,j,m = x,y,z} \chi^{(2)}_{ljm} \hat{E}_{p}^{(l) (+)}(\vec r,t)\hat{E}_{s}^{(j)(-)}(\vec r,t)\hat{E}_{i}^{(m)(-)}(\vec r,t)+ \mathrm{h.c.}.
	\label{Hpdc}
\end{equation}

In LiNbO$_3$, to close the phasematching at the telecom wavelength, an additional poling must be applied, which results in a modulation of the second order susceptibility with a distance as $\chi^{(2)}(z)= \chi^{(2)}_b \sum_n \mathrm{e}^{i \frac{2 \pi n }{\Lambda } z} $, where $\Lambda$ is the poling period and $\chi^{(2)}_b$ is the nonlinear susceptibility of the non-poled LiNbO$_3$. Taking into account only the first-order term in $\chi^{(2)}(z)$ and substituting Eqs. (\ref{signal_field}), (\ref{pump_field}) in Eq.~(\ref{Hpdc}), we obtain the following expression for the Hamiltonian:
\begin{equation}
	\hat H_\mathrm{PDC}(t)= \mathrm{i} \hbar \mathrm{G} \mathrm{O}
	\int \mathrm{d} \omega_s \mathrm{d} \omega_i \mathrm{e}^{-\frac{t^2}{2\tau^2}} \mathrm{e}^{-\mathrm{i}(\omega_p-\omega_s-\omega_i) t} \int_0^L \mathrm{d}z \mathrm{e}^{\mathrm{i}(k_p-k_s-k_i- \frac{2 \pi }{\Lambda }) z} \hat a_{s}^\dagger (\omega_s) \hat a_{i}^\dagger (\omega_i), 
	\label{Hpdc1}
\end{equation}
where $\mathrm{G}= \frac{\mathrm{i} \mathcal{E}_{p} }{12\pi c } \sqrt{\frac{\omega_{0s} \omega_{0i} n_\mathrm{eff}(\omega_{0s})n_\mathrm{eff}(\omega_{0i}) }{S_\mathrm{eff}(\omega_{0s}) S_\mathrm{eff}(\omega_{0i}) }}$ and the overlapping coefficient 
\begin{eqnarray}
	\mathrm{O} = \int \mathrm{d}x \ \mathrm{d}y \sum_{l,j,m=x,y,z} \chi^{(2)}_{ljm} E^{*(l)}_{\perp p} (x,y) E^{(j)}_{\perp s} (x,y) E^{(m)}_{\perp i} (x,y).
\end{eqnarray}

Integrating over $z$, we get:
\begin{equation}
	\hat H_\mathrm{PDC}(t)= \mathrm{i} \hbar \mathrm{G O} L
	\int \mathrm{d} \omega_s \mathrm{d} \omega_i \mathrm{e}^{-\frac{t^2}{2\tau^2}} \mathrm{e}^{-\mathrm{i}(\omega_p-\omega_s-\omega_i) t} \sinc\left[\frac{\Delta \beta L}{2}\right]\mathrm{e}^{\mathrm{i}\frac{\Delta \beta L}{2}} \hat a_{s}^\dagger (\omega_s) \hat a_{i}^\dagger (\omega_i),
	\label{Hpdc2}
\end{equation}
where $\Delta \beta = k_p-k_s-k_i- \frac{2 \pi }{\Lambda } $.

Using the first-order perturbation theory, one can obtain the final expression for the state vector of the spontaneous PDC process:
\begin{equation}
	\begin{aligned}
		|\psi_{\mathrm{PDC}}\rangle &\simeq \frac{1}{\mathrm{i} \hbar} \int \mathrm{d}t \hat H_{\mathrm{PDC}}(t) |0\rangle\\
		&=\sqrt{2 \pi} \tau L \mathrm{G O} \int \mathrm{d} \omega_s \mathrm{d} \omega_i  \mathrm{e}^{-\frac{(\omega_p-\omega_s-\omega_i)^2\tau^2}{2}} \sinc\left[\frac{\Delta \beta L}{2}\right]\mathrm{e}^{\mathrm{i}\frac{\Delta \beta L}{2}} \hat a_{s}^\dagger (\omega_s) \hat a_{i}^\dagger (\omega_i) |0\rangle.
	\end{aligned}
\end{equation}

Finally, the state vector can be written as
\begin{equation}
	|\psi_{PDC}\rangle = \Gamma \int \mathrm{d} \omega_s \mathrm{d} \omega_i F_N(\omega_s,\omega_i) \hat a_{s}^\dagger (\omega_s) \hat a_{i}^\dagger (\omega_i) |0\rangle,
\end{equation}
where $\Gamma=\sqrt{2 \pi} \tau L \mathrm{G O} N_F $ is the coupling strength which determines the number of generated photon pairs, $N_F = \sqrt{\int \mathrm{d} \omega_s \mathrm{d} \omega_i \left( \mathrm{e}^{-\frac{(\omega_p-\omega_s-\omega_i)^2\tau^2}{2}} \sinc\left[\frac{\Delta \beta L}{2}\right]\right)^2}$ is the normalization constant, and
\begin{eqnarray}
	F_N(\omega_s,\omega_i) = \frac{1}{N_F} \int \mathrm{d} \omega_s \mathrm{d} \omega_i \mathrm{e}^{-\frac{(\omega_p-\omega_s-\omega_i)^2\tau^2}{2}} \sinc\left[\frac{\Delta \beta L}{2}\right]\mathrm{e}^{\mathrm{i}\frac{\Delta \beta L}{2}} 
\end{eqnarray}
is the normalized joint spectral amplitude.

A connection between the coupling strength $\Gamma$ and the coupling coefficients calculated in Appendix A is approximately given by $|\Gamma| \approx \frac{1}{3} |\kappa_s| \tau N_F L \sqrt{P}$, where the factor $\frac{1}{3}$ accounts the fact that the parametric down-conversion process is one of three nonlinear second-order processes \cite{Boy08}, $P$ is the pump power.

\section{Refractive index of lithium niobate in LNOI films}\label{appc}
The optical properties of the LiNbO$_3$ in the LNOI films have been determined experimentally using spectroscopic ellipsometry. Here, a xenon arc lamp was used as a light source in combination with a silicon/(In,Ga)As stack photodiode in order to realize measurements in a broad spectral range of $400\,\text{nm}$ -- $1700\,\text{nm}$. To increase the precision of the resulting optical functions, the measurements were carried out at different angles of incidence ranging from 56$^\circ$ to 76$^\circ$ in 10$^\circ$ steps. Due to the low index contrast between LiNbO$_3$ and SiO$_2$, an auto-retarder was used for polarization control in order to obtain the main elements of the Mueller matrix. 

To model the optical properties of the LiNbO$_3$ films, three main contributions were taken into account: An offset to the dielectric function $\varepsilon_{re}$, a Sellmeier oscillator (i.e., a Lorentz oscillator without dissipation) and a Tauc-Lorentz oscillator \cite{Jel96}. All three contributions were introduced as a biaxial model to take into account the birefringence of the material. 
The overall dielectric function thus comes down to
\begin{eqnarray}
	\tilde{\varepsilon}_{\text{LiNbO3}}=\varepsilon_{re}+\varepsilon_{\text{Sellm.}}+\tilde{\varepsilon}_{\text{T.-L.}}
\end{eqnarray}
For the Sellmeier contribution we have:
\begin{eqnarray}
	\varepsilon_{\text{Sellm.}}(E(\text{eV}))=\frac{A_{n,\text{Sellm.}}E_{0,\text{Sellm.}}^2}{E_{0,\text{Sellm.}}^2-E^2}
\end{eqnarray}
The Tauc-Lorentz contribution is given by
\begin{eqnarray}
	\tilde{\varepsilon}_{\text{T.-L.}}=\varepsilon_{\text{re, T.L.}}+i\varepsilon_{\text{im, T.L.}}.
\end{eqnarray}
The imaginary part of the Tauc-Lorentz oscillator is given by:
\begin{eqnarray}
	\varepsilon_{\text{im, T.L.}}=\left\{
	\begin{array}{ll}
		\frac{A_{n,\text{T.L.}}E_{n,T.L.}C_{n,T.L.}(E-E_g)^2}{(E^2-E_{n,T.L.}^2)^2+C_{n,T.L.}^2E^2}\cdot\frac{1}{E} & E>E_g \\
		0 & E\le E_g
	\end{array}
	\right.
\end{eqnarray}
The real part is derived using the following integral:
\begin{eqnarray}
	\varepsilon_{\text{re, T.L.}}=\frac{2}{\pi}\,PV\!\!\int\limits_{E_g}^\infty\frac{\xi\,\varepsilon_{\text{im, T.L.}}(\xi)}{\xi^2-E^2}d\xi,
\end{eqnarray}
where $PV$ denotes Cauchy's Principal Value. 

An interactive fitting procedure resulted in the values for the ordinary and extraordinary refractive indices displayed in Table~\ref{tab:LNOI_refr}. The resulting refractive indices are shown in Fig.\,\ref{scan_f}.
\begin{table}[h!bt]
	\centering
	\begin{tabular}{c||ccccccc} \hline\hline
		Dir. & $\varepsilon_{\text{re}}$ & $A_{n,\text{Sellm.}}$ & $E_{0,\text{Sellm.}}$ & $A_{n,\text{T.L.}}$ & $E_{n,T.L.}$ & $C_{n,T.L.}$ & $E_g$\\
		&  &  & eV & eV & eV & eV & eV\\ \hline\hline
		$e.o.$& 1.5015 & 41.169 & 6.4608 & 282.99 & 5.0543 & 1.4905 & 4.2919 \\
		$o.$ & 1.3135 & 32.999 & 6.7798 & 497.17 & 4.9351 & 1.0358 & 4.3305 \\ \hline\hline
	\end{tabular}
	\caption{Parameters for the optical functions of LiNbO$_3$ films in LNOI substrates. The extraordinary direction is denoted using $e.o.$ in the table.}
	\label{tab:LNOI_refr}
\end{table}
\end{document}